\documentclass[twocolumn,showpacs,amsmath,amssymb,prb,aps]{revtex4}

\usepackage{graphicx}% Include figure files
\usepackage{dcolumn}% Align table columns on decimal point
\usepackage{bm}% bold math
\usepackage{psfrag}% symbal display

\def\d{\delta}

\begin{document}

\title{\Large\bf Coexistence of Superconductivity and Ferromagnetism
in Dilute Co-doped La$_{1.89}$Ce$_{0.11}$CuO$_{4\pm \d}$ System}

\author{K. Jin, J. Yuan, L. Zhao, H. Wu, X.Y. Qi, B.Y. Zhu, L.X. Cao, X.G. Qiu, B. Xu, X.F. Duan,  and B.R. Zhao}\email{brzhao@aphy.iphy.ac.cn}

\affiliation{National Laboratory for Superconductivity, Institute of
Physics and Beijing National Laboratory for Condensed Matter
Physics, \\ Chinese Academy of Sciences, P. O. Box 603, Beijing
100080, China}

\date{\today}

\begin{abstract}
Thin films of the optimally electron-doped $T'$-phase
superconductor La$_{1.89}$Ce$_{0.11}$CuO$_{4\pm \d}$ are
investigated by dilute Co doping, formed as
La$_{1.89}$Ce$_{0.11}$(Cu$_{1-x}$Co$_{x}$)O$_{4\pm\d}$ (LCCCO)
with $x$ = 0.01 -- 0.05. The following results are obtained for
the first time: for the whole dilute Co doping range, LCCCO thin
films show long-range ferromagnetic ordering at the temperature
range from 5 K to 300 K, which is likely due to the RKKY
interaction; in the very dilute Co doping, $x$ = 0.01 and 0.02,
the superconductivity is maitained, the system shows the
coexistence of superconductivity and ferromagnetism in the
CuO$_{2}$ plane. This may be based on the nature of the charge
carriers in electron-doped high-$T_{c}$ cuprate superconductors.

\end{abstract}


\maketitle

\pagebreak

An important progress in superconductor physics is the observation
of coexistence of superconductivity (SC) and ferromagnetism (FM)
in the spin-triplet pairing superconductors, such as
UGe$_{2}$,\cite{UGe2} ZrZn$_{2}$ \cite{ZrZn2} and
URhGe.\cite{URhGe} For high-$T_{c}$ cuprate superconductors
(HTSC), it is commonly believed that the  CuO$_{2}$ plane plays an
important role in SC. Therefore, seeking for the possibility of
coexistence of SC and FM in the CuO$_{2}$ plane is very helpful
for us to understand the interaction between SC and magnetism, and
the mechanism of SC. The RuSr$_{2}$GdCu$_{2}$O$_{8}$ (Ru-1212) is
well known for the feature of coexistence of SC and FM, but the
study shows that the SC takes place in the CuO$_{2}$ plane while
the FM forms in the RuO$_{2}$ plane.\cite{Ru1212} In the past
years, the effect of magnetic impurity doping in the CuO$_2$ plane
has been an important issue for HTSC. Many works have been done on
the substitution of a magnetic Ni-ion (Fe, Co, etc. are also
included in some works) for a Cu-ion in (La,Sr)$_{2}$CuO$_{4}$,
YBa$_{2}$Cu$_{3}$O$_{7}$, Bi$_{2}$Sr$_{2}$CaCu$_{2}$O$_{8}$, and
their related ones.\cite{Ni} For comparison, the same works have
been done on the nonmagnetic Zn-ion substitution.\cite{Zn} The
study of the impurity doping effect has provided a great deal of
information about the nature of the cuprates. However, there have
been no reports on the possible coexistence of SC and FM in the
CuO$_{2}$ plane.

It should be particularly pointed out that most of the previous
experiments on the magnetic impurity doping effect are focused on
the hole-doped HTSC, and only a few works have been addressed on
the electron-doped ones.\cite{etype} This mainly attributed to the
relatively lower $T_{c}$ ($\leq$ 30 K) and more complicated
preparation process of the electron-doped HTSC. However, the
electron-doped HTSC have their own intriguing properties,
especially the two-band model and two types of charge carriers
(electrons and holes) are confirmed experimentally and
theoretically.\cite{wang,jiang,arpes} It will be interesting and
challenging for us to deeply explore the interaction of SC and FM
related the CuO$_{2}$ plane by magnetic impurity doping in the
electron-doped HTSC, since even for the case of Ru-1212, the
two-band model and two kinds of charge carriers are considered to
be the origin of the coexistence of SC and FM.\cite{coexist}

In the present work, thin films of the optimally electron-doped
$T'$-phase superconductor La$_{1.89}$Ce$_{0.11}$CuO$_{4\pm \d}$
(LCCO) are investigated by dilute Co doping at the Cu site.
Through the systematic transport and magnetization measurements of
La$_{1.89}$Ce$_{0.11}$(Cu$_{1-x}$Co$_{x}$)O$_{4\pm\d}$ (LCCCO)
thin films with $x$ = 0.01 -- 0.05, together with reduction
treatment and valence examinations, some interesting new effects
are found contrasting to the previous works: the long-range FM
ordering is observed in the temperature range of 5 K -- 300 K  for
all the designed Co doping concentrations, $x$ = 0.01 to 0.05,
which is suggested to be formed by the RKKY interaction; for the
cases of very dilute Co concentrations, $x$ = 0.01 and 0.02, the
coexistence of SC and FM at the temperatures below $T_{c}$ in the
CuO$_{2}$ plane is obviously detected; the electrons play an
important role for SC in LCCO system.

\begin {figure}[!]  %\graph1
\begin{center}
\includegraphics[bb=18 18 591 305, width=8cm, clip]{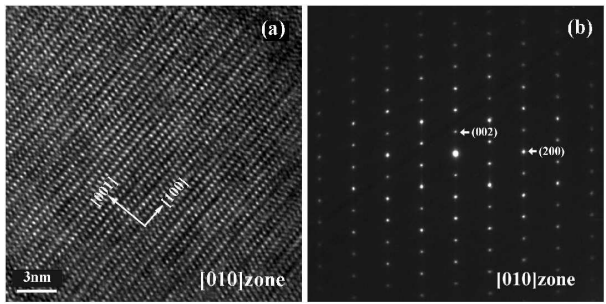} %jpg%
\caption{\label{hrem} The [010] zone axis HREM image (a) and its
corresponding electron diffraction pattern (b) of LCCCO sample
with Co concentration $x = 0.05$.} %\href{*}{*}
\end{center}
\end{figure}

\begin {figure}[!]  %\graph2
\begin{center}
\includegraphics*[bb=19 19 591 296, width=8cm, clip]{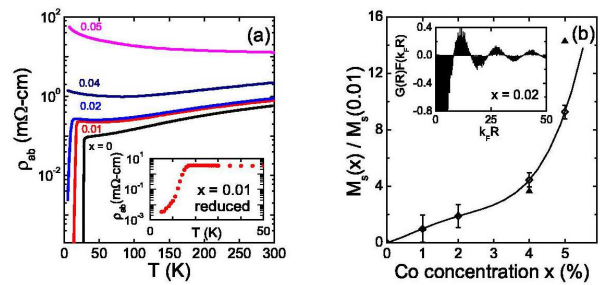} %jpg
\caption{\label{rh} (Color online) (a) In-plane resistivity
$\rho_{ab}$ versus temperature for the films with Co concentration
$x$ = 0, 0.01, 0.02, 0.04, 0.05. Inset: $\rho_{ab}(T)$ of the
reduction-treated film (after 3 times annealing) with $x$ = 0.01.
(b) $M_{S}(x)$ obtained by 2D damped RKKY model characterization
(normalized by $M_{S}(x = 0.01)$ with $l$ = 50 $\rm{\AA}$, $k_{F}$
= 0.8 $\rm{\AA^{-1}}$. Triangles and diamonds represent
experimental data and calculated value, respectively. Solid line
is the fit for the data. The error bars are from the statistical
errors for 100 times calculations. Inset: typical oscillating
behavior of $G(R)F(k_{F}R)$ versus $k_{F}R$ for the thin film with
$x$ = 0.02. }
%\href{}{}
\end{center}
\end{figure}

The optimally doped LCCO is used as the benchmark material for the
present Co doping study, since its normal state is in a completely
metallic phase and it has the highest $T_{c}$ ($\sim$ 30 K) in the
electron-doped HTSC family,\cite{lcco} we can easily test the
influence of Co substitution for Cu \cite{site} on SC at
relatively high temperatures. The condition of preparation of
LCCCO thin films is similar to the case of LCCO ones as described
in detail elsewhere.\cite{wu,zhao} The stoichiometric targets with
the atomic ratio of La:Ce:Cu:Co = 1.89:0.11:(1-$x$):$x$ were
fabricated by a solid state reaction process. Then LCCCO thin
films with $x$ = 0.01, 0.02, 0.04 and 0.05 were deposited on
(001)-oriented SrTiO$_{3}$ (STO) substrates by dc magnetron
sputtering with $\sim$ 250 $nm$ thickness each. X-ray diffraction
(XRD) data show that all the thin films are in single $T'$-phase
and (001)-oriented. Fig.~\ref{hrem} shows the [010] zone axis high
resolution electron microscopy (HREM) image and the corresponding
electron diffraction pattern of the film with $x$ = 0.05, which
indicates a perfect structure. In order to examine the electric
and magnetic properties, the in-plane resistivity and
magnetization were measured. All measurements were carried out
using a Quantum Design MPMS-5 equipment. In order to measure the
resistivity, the samples were patterned into the  bridges with
width of 50 $\mu m$. Then the Ag electrodes were deposited on the
surfaces of the films through a metal mask.
The in-plane resistivity $\rho_{ab}(T)$ data as shown in
Fig.~\ref{rh} (a) indicate that the samples with $x$ = 0.01, 0.02
are superconducting, with zero resistance temperature $T_{c0}$
$\sim$ 13 K and $<$ 5 K, respectively. $T_{c0}$ is $\sim$ 27 K for
the pure LCCO thin film (with $x$ = 0). While the samples with $x
=$ 0.04 and 0.05 tend to be insulator-like with decreasing $T$
down to 5 K. The magnetization versus temperature $M(T)$ curves,
with magnetic field $H$ = 1000 Oe (parallel to the CuO$_{2}$
plane), show the evolution from superconductor to ferromagnet with
increasing Co concentration. The magnetization $M(H)$ measurements
are also done for all the designed Co-doped thin films, and the
clear hysteresis loops indicate that a real FM long-range ordering
rather than other magnetic phase \cite{super} forms even for the
very dilute Co concentration, $x$ = 0.01. With increasing Co
concentration, the FM ordering enhances. It can be clearly
observed in Fig.~\ref{rh} (b) that the saturation magnetization
$M_{S}$ for the film with $x$ = 0.05 is more than 10 times larger
than that of the film with $x$ = 0.01.

\begin {figure}[!]  %\graph3
\begin{center}
\includegraphics*[bb=14 14 274 294, width=8cm,clip]{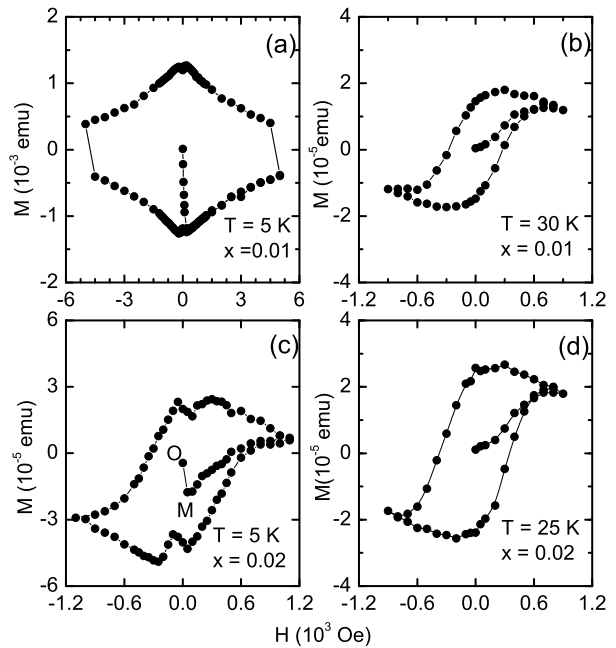}
\caption{\label{mh} In-plane Magnetization versus field for LCCCO
thin films by zero magnetic field cooling: (a) superconducting
magnetization behavior of the film with $x$ = 0.01 at 5 K; (b)
full FM hysteresis loop of the film with $x$ = 0.01 at 30 K; (c)
clear evidence for the coexistence of SC and FM of the film with
$x = 0.02$ at 5 K, confirmed by the shielding effect in the lower
field (indicated by OM, with $H_{c1} \sim 50$ Oe), and the FM
hysteresis develops with increasing $H$; (d) full FM hysteresis
loop for the film with $x = 0.02$ at 25 K. All these clearly show
the coexistence and competition of SC and FM for the LCCCO films
with $x$ = 0.01 and 0.02.}
%\href{}{}
\end{center}
\end{figure}

Surprisingly, $M(H)$ data for the samples with $x$ = 0.01 and 0.02
definitely show the coexistence of SC and FM in the
superconducting transition region. In Fig.~\ref{mh}, four typical
$M(H)$ curves are presented: (a) a superconducting $M(H)$ curve
for the sample with $x = 0.01$ is observed at the temperature (T =
5 K) lower than $T_{c0}$; (b) the full ferromagnetic hysteresis
loop for the same sample at 30 K (higher than $T_{c}$); (c) the
$M(H)$ data of the sample with $x$ = 0.02 give clearly evidence
for the shielding effect in low field region (indicated by OM,
with $H_{c1}$ of $\sim$ 50 Oe at 5 K). However, with increasing
$H$, SC is suppressed and FM becomes dominant in the sample; (d)
with increasing the temperature up to 25 K (higher than $T_{c0}$),
the curve for the same sample with $x$ = 0.02 is transformed into
a full FM hysteresis loop. These four loops obviously reveal the
coexistence and competition of SC and FM in the very dilute Co
concentrations.

For the origin of FM ordering in LCCCO, several possibilities
should be considered.\cite{apl} One is the segregated Co clusters
that can be formed if the Co atoms can not dissolve into the
lattice of the LCCO.
If such dilute Co atoms gather to form FM, the large area of the
film must be Co-ion-free, which results in seperation of SC basis
with a few FM clusters, and then no above regular influence of Co
doping on SC can be observed. Furthermore, We grew (103)-oriented
LCCCO thin films on the miscut (110)-oriented STO substrates in
the same batch with the (001)-oriented sample. The process is
similar to the growth of (103)-oriented LCCO thin films.\cite{wu}
The magnetization data show that the $M_{S}$ in (103) plane is
$\sim$ $1.4\times10^{-5}$ emu, obviously smaller than the $M_{S}$
$\sim$ $1.8\times10^{-5}$ emu in (001) plane. Such anisotropy of
$M_{S}$ strongly suggestes that the FM ordering should not
originate from the segregated Co clusters. Actually, no Co cluster
is observed in other Co-doped systems, such as the Co-doped
SnO$_{2}$ \cite{sno} and ZnO,\cite{apl} in which the long-range FM
ordering with the Curie temperature higher than room temperature
is formed, while no Co cluster is observed. Another possible
origin of the FM ordering, i.e., the Co oxides, can also be
excluded, because almost all of the Co oxides are antiferromagetic
or ferrimagnetic.

Therefore, the Ruderman-Kittle-Kasuya-Yosida (RKKY) interaction,
i.e., the coupling of the localized magnetic moments via polarized
charge carriers, should be the dominant mechanism for the FM
ordering in the present LCCCO system. Here, a 2D damped RKKY
interaction model \cite{408} is used to characterize the
experimental data. The RKKY interaction can be given as the form,
\cite{408,2d}
\begin{eqnarray}
H=\frac{J^{2}V^{2}m^{*}k_{F}^{2}}{8\pi^{2}\hbar^{2}N^{2}}
\sum_{\scriptstyle ij \atop \scriptstyle i\neq j}
\exp{(-\frac{R_{ij}}{l})}F(k_{F}R_{ij}){\bm S}_{i}\cdot{\bm
S}_{j}\label{eq},\\ \nonumber\mbox{with\qquad\qquad\qquad}
F(x)\equiv J_{0}(x)N_{0}(x)+J_{1}(x)N_{1}(x),
\end{eqnarray}
where $J$, $m^{*}$, $k_{F}$, $S_{i}$, and $R_{ij}$ are
respectively the interaction between the local magnetic moment and
the polarized carrier, the effective mass of the conduction
carrier, the Fermi wave vector, the $i$th Co local moment, and the
distance between $S_{i}$ and $S_{j}$ moments; V and N denote the
volume and the cell numbers of the samples, respectively; $l$ is
the mean free path of the carriers; $J_{i}$ and $N_{i}$ ($i$ = 0,
1) are the first and the second Bessel functions, respectively.
Employing a simplified Monte Carlo method,\cite{apl} we can
rewrite Eq.~(\ref{eq}) as $H_{RKKY} = C(S\cdot
S)\sum_{R}\exp{(-R/l)}G(R)F(k_{F}R)$, with constant $C$ and $S$
(spin), the Co-ion distribution function $G(R)$, and the
oscillating function $F(k_{F}R)$. If the sum of Eq.~(\ref{eq}) is
negative, the FM can be achieved favorably. For each Co doping
case, we make calculations in a 2D 2500 $\times$ 2500 lattice and
average in 100 times. Since this sum is not sensitive to $k_{F}$
so long as assuring it is negative (FM forms), it is reasonable to
choose $k_{F}$ = 0.8 $\rm{\AA^{-1}}$. We set $l$ = 50 $\rm{\AA}$
in order to keep the carriers moving well. Then the reduced
$M_{s}$ for each Co concentration is obtained from the sum of
$\exp{(-R/l)}G(R)F(k_{F}R)$,\cite{2d,kan} which is reduced by
$M_{S}(x = 0.01)$ as shown in Fig.~\ref{rh}(b). The calculations
of $M_{s}$ are well in agreement with the experiment data except
for the sample with $x = 0.05$, where a large bias between the
theoretical calculation and the experimental data may be caused by
the crossover of FM ordering from 2D to 3D. The inset of
Fig.~\ref{rh}(b) shows the oscillating behavior of the RKKY
interaction for $x = 0.02$.
This calculation clearly indicates that with increasing Co
concentration, the RKKY interaction is enhanced due to the
decrease of distances between Co-ion moments. It also indicates
that the RKKY interaction be the dominant mechanism for the
long-range FM ordering. Comparing with other magnetic impurity
doping, such strong RKKY interaction may be originated from the
feature of Co, e.g., strong s-d electrons exchange interaction and
high Curie temperature. Such strong RKKY interaction has also been
observed in other Co-doped compounds.\cite{apl} The RKKY
interaction needs the charge carriers in the CuO$_{2}$ plane, and
the charge carriers must also take responsibility for the SC in
the superconducting state. How to solve such two roles of charge
carriers in the CuO$_{2}$ plane? We consider that the existence of
two types of charge carriers in electron-doped HTSC
\cite{wang,jiang,arpes} may be important. The previous work
\cite{coexist} has suggested that the two-band model and two kinds
of charge carriers in Ru-1212 take responsibility for the
existence of SC and FM. Of course, understanding the origin of
such coexistence of SC and FM in LCCCO should be probed further.

\begin {figure}[!]  %\graph4
\begin{center}
\includegraphics*[bb=45 237 558 717, width=8cm, clip]{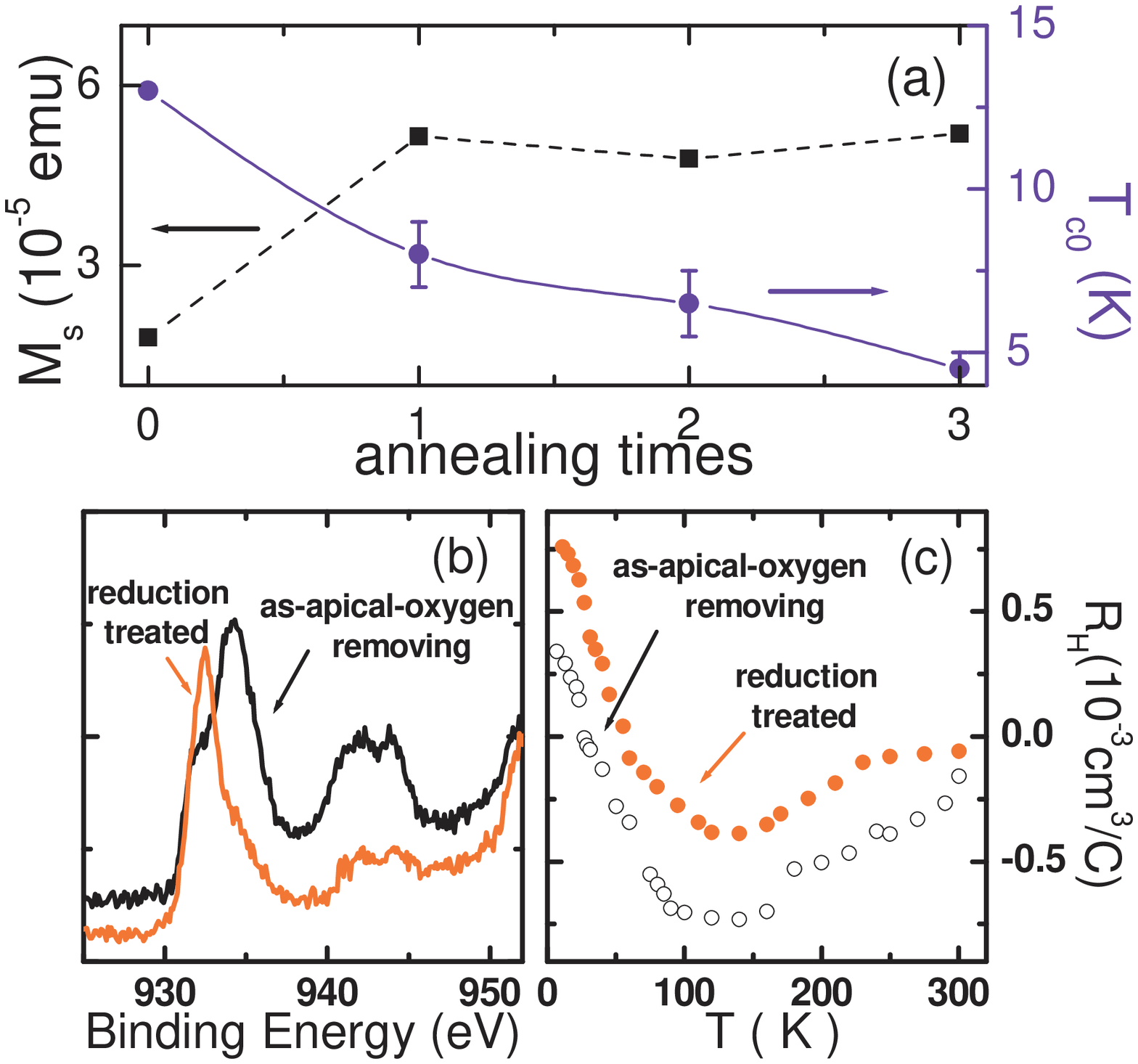}
\caption{\label{rkky} (color online) (a) $M_{S}$ (square) and
$T_{c0}$ (circle) versus annealing times, the reduction treatment
enhances FM to high level, whereas $T_{c0}$ is decreased
remarkably, which is caused by the reduction of the valence of Cu.
(b) and (c) Cu $2p_{3/2}$ core-level XPS spectra and Hall
coefficient of the cases of as-apical-oxygen removing and
reduction-treated LCCCO with $x$ = 0.01, respectively.}
%\href{}{}
\end{center}
\end{figure}

In order to understand the possible role of charge carriers on SC
and FM ordering further, a reduction treatment is carried out for
the sample with $x = 0.01$, through which the relative strength of
SC and FM will be changed by adjusting the concentrations of
charge carriers. The samples with $x = 0.01$ were annealed at
$\sim$ 600 K in $\sim$ $10^{-5}$ Pa vacuum for several minutes.
The key for this process is to partially remove the oxygen from
the films, but maintain the single $T'$-phase. After the
treatment, XRD data show that the films are still single
$T'$-phase. Fig.~\ref{rkky}(a) shows the evolution of the $M_{S}$
with the annealing times. It shows that at 15 K the $M_{S}$ is
about 2.8 times larger after the first time annealing in vacuum,
and then keeps almost constant for the further annealing steps,
whereas $T_{c0}$ decreases from 13 K before reduction treatment
(the case of as-apical-oxygen removing) to $<$ 5 K after the third
time annealing.

In order to reveal the origin of the changes of SC and FM with the
reduction treatment, the X-ray Photoelectron Spectroscopy (XPS)
measurements were performed with an AXIS-Ultra instrument from
Kratos. The valences of La, Ce, Co and Cu were tested before and
after the reduction treatment. The results clearly show that after
reduction treatment, the main peak of the Cu $2p_{3/2}$ state
changes from 934.3 eV to 932.5 eV, and the ratio of the Cu
$2p_{3/2}$ satellite peak to Cu $2p_{3/2}$ main peak decreases as
shown in Fig.~\ref{rkky}(b). That is, the valence of partial Cu
ions changes from +2 to +1.\cite{xps} While the valences of La, Ce
and Co have no detectable changes. The valence change of Cu
presents that the reduction process preferentially removes the
oxygen O(1) from the CuO$_{2}$ plane, rather than the O(2) in La-O
plane, which is similar to that induced by the reduction process
in NCCO ,\cite{osite} where the oxygen was removed first from the
CuO$_{2}$ plane with the reduction treatment. Owing to the
reduction of Cu ions, this process should result in the decrease
of the concentration of the electrons in the CuO$_{2}$ plane and
weakens the SC. However, the RKKY interaction is enhanced ($M_{S}$
increases) after reduction treatment, so the concentration of
holes possibly increases (the RKKY interaction needs the charge
carriers). From the Hall measurements as shown in
Fig.~\ref{rkky}(c), we find that the Hall coefficient is shifted
towards positive direction after reduction treatment indeed.
Therefore, the decrease of $T_{c}$ should be attributed to the
reduction of the concentration of electrons. This means that the
electrons may play an important role for SC in the LCCO, which
seems to be different from the point of view that the holes are
more important to SC in the NCCO.\cite{hole}

In conclusion, the optimally doped LCCO thin films are
investigated by dilute magnetic impurity Co doping at the Cu site.
It is observed for the first time that the long-range FM ordering
appears at the temperature range from 5 K to 300 K for all the
LCCCO thin films, $x$ = 0.01 -- 0.05, which is suggested to be
formed by the RKKY interaction; the coexistence of SC and FM is
observed in the very dilute Co concentrations, $x$ = 0.01 and
0.02, below $T_{c}$. The existence of two kinds of charge carriers
in the electron-doped HTSC is suggested to be an important reason
to such coexistence. However, other reason, such as that the
Co-ion may have some special feature when it is doped in
electron-doped HTSC, should also be considered. Based on the
reduction experiment, we argue that the electrons play an
important role for SC in LCCO system. We believe that the present
work provides new information for understanding the intrinsic
feature of the electron-doped HTSC.

We thank Prof. F.C. Zhang, Prof. X.C. Xie, Prof. P.C. Dai, Prof.
D.S. Wang and Prof. W.M. Liu for fruitful discussions. This work
is supported by grants from the State Key Program for Basic
Research of China and the National Natural Science Foundation.

\end{document}